\documentclass[runningheads]{llncs}
\usepackage{graphicx}
\usepackage{cite}
\usepackage{amsmath,amssymb,amsfonts}
\usepackage{algorithmic}
\usepackage{graphicx}
\usepackage{xcolor}
\usepackage{soul}
\usepackage{diagbox}
\usepackage[normalem]{ulem}
\useunder{\uline}{\ul}{}
\usepackage{mathtools}

\usepackage{url}
\usepackage{paralist}
\usepackage{upgreek}
\usepackage{subcaption}
\usepackage{multirow}
\usepackage{booktabs}
\usepackage{colortbl}

\renewcommand{\figurename}{Figure}
\newcommand{\etal}{~\textit{et~al.}}

\begin{document}
\title{SAMedOCT: Adapting Segment Anything Model (SAM) for Retinal OCT}

\author{Botond Fazekas \and
Jos\'e Morano \and
Dmitrii Lachinov \and
Guilherme Aresta \and
Hrvoje Bogunovi\'c}

\institute{Christian Doppler Laboratory for Artificial Intelligence in Retina, Department of Ophthalmology and Optometry, Medical University of Vienna, Austria\\\email{\{botond.fazekas,hrvoje.bogunovic\}@meduniwien.ac.at}}

\authorrunning{B. Fazekas\etal}

\titlerunning{SAMedOCT: Adapting Segment Anything Model (SAM) for Retinal OCT}

\maketitle              
\begin{abstract}
The Segment Anything Model (SAM) has gained significant attention in the field of image segmentation due to its impressive capabilities and prompt-based interface. While SAM has already been extensively evaluated in various domains, its adaptation to retinal OCT scans remains  unexplored. To bridge this research gap, we conduct a comprehensive evaluation of SAM and its adaptations on a large-scale public dataset of OCTs from RETOUCH challenge. Our evaluation covers diverse retinal diseases, fluid compartments, and device vendors, comparing SAM against state-of-the-art retinal fluid segmentation methods. Through our analysis, we showcase adapted SAM's efficacy as a powerful segmentation model in retinal OCT scans, although still lagging behind established methods in some circumstances. The findings highlight SAM's adaptability and robustness, showcasing its utility as a valuable tool in retinal OCT image analysis and paving the way for further advancements in this domain.
\end{abstract}
\section{Introduction}
Optical coherence tomography (OCT) has become the gold-standard imaging modality in ophthalmology, in particular for retinal diseases affecting the macula and consequently the central vision, such as age-related macular degeneration (AMD), retinal vein occlusion (RVO), and diabetic macular edema (DME). Its ability to provide three-dimensional cross-sectional views of the retina in a fast and non-invasive manner has proven invaluable in the management of patients with neovascular AMD, RVO, and DME, characterized by the onset of fluid into the macula, and in guiding their treatment with anti-VEGF drugs~\cite{CAMPOCHIARO2016S78}. However, in the current clinical practice, the clinical assessment of retinal fluid is still primarily qualitative, which is subjective, inaccurate, and time-consuming. 

Deep learning has revolutionized the field of medical image analysis, with a high potential for empowering clinicians with robust second opinions, quantitative measures of biomarkers via image segmentation, and even (semi-)automation of the patient management workflow. 
The accurate segmentation of healthy and pathological biomarkers allows for quantitative analysis of these images, easing early disease detection, monitoring, and patient follow-up. This made the self-configuring nnU-net~\cite{isensee_2021} model especially popular as a go-to solution for training segmentation models for tasks in medical and ophthalmic imaging. However, the variety and complexity of the retinal anatomical structures and the low signal-to-noise ratio of OCT make this task still very challenging.

Recently, the deep learning community has focused on the development of foundation models~\cite{moor_2023}. These models leverage large-scale pre-training on massive datasets, enabling them to learn rich representations and capture intricate visual patterns. While the majority of such models were developed for the purpose of image interpretation, few targeted image segmentation task, likely the most important one for ophthalmic image analysis. This changed when Segment Anything Model (SAM) was introduced in April 2023~\cite{kirillov_2023}. Trained on 11 million images and 1 billion masks, it demonstrated excellent capability in semantic segmentation of natural images, including \emph{zero-shot} generalization. However, the potential of SAM for medical image segmentation is still unclear and is being extensively explored~\cite{zhang_2023}. Specifically, SAM potential for retinal OCT segmentation, and how best to exploit it for this task is currently unknown.

In this paper, we examine the performance of SAM-based models for retinal OCT segmentation, for the purpose of measuring fluid volumes in an automated manner, the most important biomarkers for the management of patients with macular edema. We perform a large-scale evaluation of different SAM modalities on a public dataset originating from a MICCAI challenge \cite{2019_Bogunovic}.
We first analyze different SAM modalities, from a zero-shot setup to multi-click and box selection. Then, more importantly, we show that SAM can be effectively adapted and trained to become a powerful segmentation model in retinal OCT.

\begin{figure}[tbh!]
  \centering

  \begin{subfigure}{0.3\linewidth}
    \includegraphics[width=\linewidth]{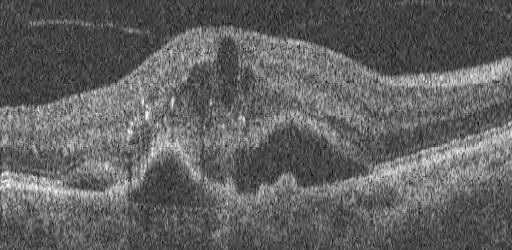}
    \caption{Original B-Scan}
    \label{fig:sam_original}
  \end{subfigure}%
  \hfill
  \begin{subfigure}{0.3\linewidth}
    \includegraphics[width=\linewidth]{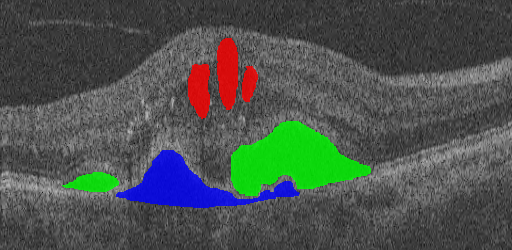}
    \caption{Groundtruth}
    \label{fig:sam_gt}
  \end{subfigure}%
  \hfill
  \begin{subfigure}{0.3\linewidth}
    \includegraphics[width=\linewidth]{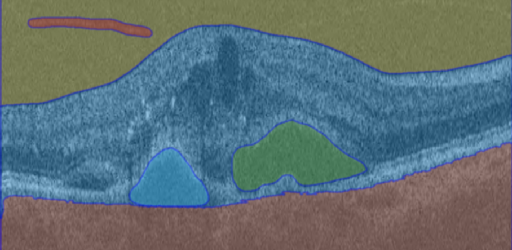}
    \caption{SAM \emph{everything} mode}
    \label{fig:sam_everything}
  \end{subfigure}

  \vspace{0.1em}

  \begin{subfigure}{0.3\linewidth}
    \includegraphics[width=\linewidth]{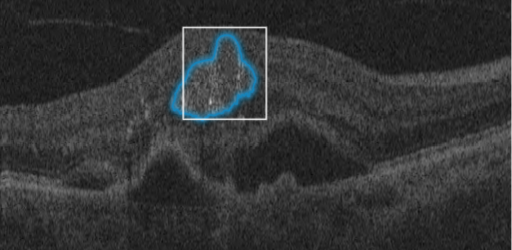}
    \caption{Box selection}
    \label{fig:sam_box}
  \end{subfigure}%
  \hfill
  \begin{subfigure}{0.3\linewidth}
    \includegraphics[width=\linewidth]{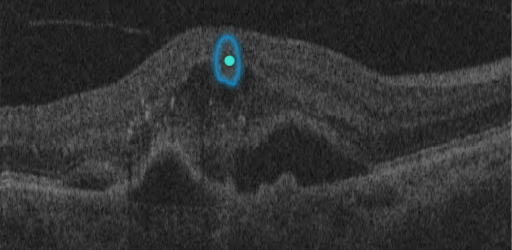}
    \caption{1 point}
    \label{fig:sam_one_click}
  \end{subfigure}%
  \hfill
  \begin{subfigure}{0.3\linewidth}
    \includegraphics[width=\linewidth]{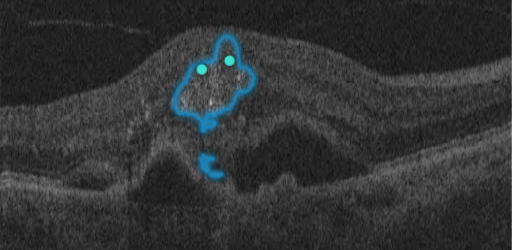}
    \caption{2 points}
    \label{fig:sam_two_clicks}
  \end{subfigure}

  \caption{An overview of the SAM segmentation modalities. In \emph{everything} mode, the model returns the masks deemed by itself as the most important. By prompting it with a bounding box, the model reduces the search space within the box and returns the best-fitting mask from it. The model can be also prompted by providing positive positions (e.g. belonging to a region of interest (ROI)), or negative positions (not belonging to the ROI)}
  \label{fig:sam_modalities}
\end{figure}

\section{Background and related works}



\subsubsection{SAM model.}
SAM is composed of three main components: an image encoder, a prompt encoder, and a mask decoder.
The image encoder is a Vision Transformer (ViT)~\cite{dosovitskiy2020image}
that takes one image as input and outputs the image embedding.
The prompt encoder has three different branches, one for each prompt type.
The branch for mask prompts consists of a simple CNN;
for points/boxes (Fig.~\ref{fig:sam_modalities}), of a lightweight embedding module; and for text, of the text encoder of CLIP~\cite{radford2021learning}.
The mask decoder takes the embeddings from the image encoder and the prompt encoder as input and outputs the final segmentation mask.
It consists of two modified Transformer decoder blocks followed by a segmentation head.

\subsubsection{SAM for medical image segmentation.}
Excelling at natural image segmentation, SAM
provides a promising foundation for expert-level methods for analyzing medical data. However, its straightforward application encounters a number of challenges specific to medical image analysis. As discussed by Zhang\etal~\cite{zhang_2023}, structural complexity, low contrast and inter-observer variability limits zero-shot application of SAM.
Furthermore, previous works~\cite{ji2023sam,huang2023segment,deng2023segment,ji2023segment,roy2023midl} confirm unsatisfactory performance for the majority of medical images, high dependence on the human prior knowledge (\textit{i.e.}, the quality of the points/boxes) and unstable performance across different datasets.
For these reasons, recent works have been focusing on fine-tuning SAM for fully automatic medical image segmentation.

Of the three components of SAM, most methods focus on 
adapting and fine-tuning the prompt encoder and the mask decoder, while keeping the image encoder frozen.
Ma\etal~\cite{ma2023segment} propose fine-tuning the mask decoder on a large set of multimodal medical images. 
Utilizing only 
box prompts, the method outperformed SAM
by a large margin on
3D and 2D tasks. Although their fine-tuned models do not
reach
the level of specialized methods, the results highlight the potential of SAM
for medical image analysis.
Similarly,
Hu\etal~\cite{hu_2023}
propose to freeze SAM image encoder and train a lightweight task-specific prediction head, ignoring the prompt encoder. The method achieves 
promising results in a few-shot learning scenario.
However, on a large
dataset, the method proved to be inferior to U-Net~\cite{isensee_2021}.
%
SAMed~\cite{zhang_2023a}
proposes to fine-tune the prompt encoder, its default input embedding, and the mask decoder of SAM, and apply a low-rank-based fine-tuning strategy (LoRA)~\cite{hu2021lora} to the image encoder.
Thus, only lightweight LoRA layers are fine-tuned, while the rest of the image encoder is kept frozen, highly reducing the training cost.
SAMed comes close to the state of the art on Synapse multi-organ segmentation dataset.
A similar approach is proposed by Wu\etal~\cite{wu_2023}.
Following the popular NLP Adapter method~\cite{houlsby2019parameter}, the authors propose to insert adapter modules at specific locations of the image encoder and the mask decoder of SAM.
Moreover, the attention operation is split into two branches as 2D space + 1D depth to account for 3D images.
Adapters are pre-trained using a self-supervised approach, and
the model is trained using a combination of point and text prompts. The
approach demonstrated competitive results on various medical tasks.
%
DeSAM~\cite{gao_2023} proposes
to fine-tune
two modules added on top of the frozen
image encoder and prompt encoder, to which
random points or a full-size box are fed.
The first module is a Transformer-based decoder that takes the outputs of both encoders and computes its embedding.
Then, the second module, a U-Net-like decoder, processes this embedding and the features from the image encoder at different scales to produce the final segmentation. 
Authors reported that DeSAM surpassed other state-of-the-art methods on a cross-site private dataset.
%
An extension of SAM to the domain of 3D medical images has been introduced by Lei\etal~\cite{lei_2023}.
The method proposes a few-shot localization framework, MedLAM, for detecting 3D anatomical regions. 
The slices within selected region are then processed with SAM or MedSAM~\cite{ma2023segment}.
The method achieves moderate performance on a range of datasets.

\section{Methods}

Given an input image $\mathbf{x} \in \mathbb{R}^{H \times W}$, where $H$ and $W$ are the height and width of the image, respectively, we aim to obtain a segmentation mask $\mathbf{y} \in \mathbb{R}^{H \times W \times C}$, where $C$ is the number of classes, using SAM-based models.
In particular, we have chosen to focus on SAMed~\cite{zhang_2023a} for several reasons.

One key factor driving our decision is the lightweight image encoder fine-tuning solution of SAMed, which offers an efficient approach to adapt SAM for medical image segmentation. This allows us to fine-tune the model while keeping computational costs and storage requirements at a minimum.


Additionally, our decision was influenced by the limitations we encountered when attempting to achieve state-of-the-art results using some of the alternative models mentioned in the related works section. Despite our efforts, these models did not yield the desired performance levels on the retinal OCT fluid segmentation task we were addressing, or required extensive unsupervised pre-training~\cite{hu_2023,wu_2023}. Considering these factors, SAMed emerged as a promising adaptation method that addressed both the efficiency requirements and the need for improved performance. In the subsequent sections, we discuss the specific modifications introduced by SAMed and our training strategy for training its OCT adapted version, SAMedOCT.


\subsubsection{SAM with LoRa}
SAMed applies the LoRA technique~\cite{hu2021lora} to the query and value projection layers of each transformer block in the image encoder of SAM. This technique serves as a bypass to achieve low-rank approximation in these layers. The authors of SAMed observed that applying LoRA only to these specific layers leads to improved performance.

To enable fast and automatic medical diagnosis during inference, SAMed eliminates the need for prompts. The default embedding utilized by the prompt encoder in SAM when no prompt is provided is retained and made trainable during the fine-tuning process. This ensures that the prompt encoder adapts and learns from the specific medical image segmentation tasks, enhancing its performance in handling different input scenarios.

SAMed introduces slight modifications to the segmentation head of the mask decoder in SAM. This modification customizes the output for each segmented semantic class, unlike the ambiguity prediction of SAM. SAMed predicts each semantic class of interest and the background in a deterministic manner, improving the interpretability and specificity of the segmentation results.

\subsubsection{Training strategy}
SAMedOCT utilizes both cross-entropy and dice losses to supervise the fine-tuning process. Similar to SAM, these losses are applied to the downsampled ground truth, as the output of SAMedOCT has a lower spatial resolution compared to the input. During our experiments, we followed the training strategy suggested in SAMed, which involves using the AdamW optimizer with a warmup period, followed by an exponential learning rate decay.

\section{Experimental setup}

To evaluate the enhancement achieved by the LoRA adaptation of the SAM model for retinal OCT scans, a comparative analysis was conducted against several baseline methods, including the original SAM model with point prompts, the SAM model with fine-tuned decoder, and state-of-the-art methods, including the methods participating in the MICCAI 2017 RETOUCH challenge~\cite{2019_Bogunovic}, and the nnU-Net model~\cite{isensee_2021}.

\subsubsection{Zero-shot SAM with point prompts} To simulate the point prompts for the SAM model, centroids were computed for each connected component of the manual reference segmentation masks. For the simulation of $n$ clicks per fluid class, the centroids of the $n$ largest connected components were utilized. In cases where the total number of connected components was smaller than $n$, a random connected component was selected, and a random coordinate was generated from a 2D Gaussian distribution centered at the chosen component's centroid. The random selection process was repeated if the generated point fell outside the mask boundaries.

\subsubsection{SAM decoder fine-tuning} The SAM model was trained with the same settings as the SAMedOCT model, but without incorporating the LoRA adaptations. The decoder component of the model was refined, while the encoder weights were kept frozen during the training process.

\subsubsection{nnU-net} We compared the SAMedOCT against the current state-of-the-art on the RETOUCH dataset, the nnU-Net model \cite{2021_Isensee}, as it was described in \cite{2023_Ndipenoch}.

\subsection{Dataset and evaluation}

All methods were trained and tested on the public RETOUCH dataset~\cite{2019_Bogunovic}. The dataset consists of 112 macula-centered OCT volumes from 112 patients with macular edema secondary to AMD or to RVO. The training set consists of 70 OCT volumes and the test set of 42. There is an approx. equal number of OCT volumes acquired with each of three OCT devices: Cirrus HD-OCT (Zeiss Meditec), Spectralis (Heidelberg Engineering), and T-1000/T-2000 (Topcon). All volumes cover a macular area of 6$\times$6 mm$^2$.

In RETOUCH, each B-scan was manually annotated pixel-wise (Fig.~\ref{fig:sam_gt}) for intraretinal fluid (IRF), subretinal fluid (SRF), and pigment epithelial detachment (PED). The test set contained double annotations coming from two medical centers and only the pixels with consensus among the two annotations were used in the evaluation following the RETOUCH challenge protocol to facilitate the comparison with the results of the original challenge. In accordance with the challenge, two metrics were utilized for the evaluations:
\begin{inparaenum}
    \item \emph{Dice coefficient}: to quantify the voxel overlap between the prediction (X) and the manual reference (Y). and
    \item \emph{Absolute volume difference} (AVD) in mm$^3$: This metric provides a clinically significant parameter by measuring the absolute difference between the volumes of X and Y: $\text{AVD} = \text{abs}(|X| - |Y|)$
\end{inparaenum}

\subsubsection{Training details}
The training was carried out in a mixed-precision environment with an Nvidia A100 (80GB) GPU, in a Singularity \cite{2017_Kurtzer} 3.7.3 environment based on the \emph{pytorch:2.0.1-cuda11.7-cudnn8-devel} Docker image, with Python 3.10.11 and Pytorch 2.0.1. The training of SAMedOCT took 16 hours on this setup, while the training of the decoder-only configuration took 4 hours.

\section{Results}
\begin{table}[b]
\centering
\setlength{\tabcolsep}{10pt}

\resizebox{\linewidth}{!}{%
\begin{tabular}{l|ccc|ccc} 
\toprule
\multicolumn{1}{l}{\multirow{2}{*}{\textbf{Experiment}}} & \multicolumn{3}{c}{\textbf{Dice} $\uparrow$}                                      & \multicolumn{3}{c}{\textbf{AVD} $\downarrow$}                     \\
\multicolumn{1}{l}{}                                     & \textbf{IRF}     & \textbf{SRF}     & \multicolumn{1}{c}{\textbf{PED}} & \textbf{IRF}    & \textbf{SRF}    & \textbf{PED}     \\ 
\midrule
SAM with 1 point                                         & $0.209$  & $0.168$  & $0.098$                  & $0.397$  & $0.676$  & $3.196$   \\
SAM with 3 points                                        & $0.260$  & $0.184$  & $0.148$                  & $0.650$  & $0.821$  & $2.526$   \\
SAM with 10 points                                       & $0.402$ & $0.406$ & $0.480$                 & $0.430$ & $0.144$ & $0.255$  \\ 
\midrule
SAM with fine-tuned decoder                                        & $0.627$ & $0.286$ & $0.448$                 & $0.055$ & $0.117$ & $0.182$  \\ 
\midrule
SAMedOCT                                                    & $\mathbf{0.766}$   & $\mathbf{0.759}$   & $\mathbf{0.815}$                   & $\mathbf{0.042}$   & $\mathbf{0.020}$   & $\mathbf{0.033}$    \\
\bottomrule
\end{tabular}
}
\vspace{5pt}
\caption{Experiments for evaluating the segmentation performance of different SAM settings.}
\label{tab:ablation}
\end{table}

\begin{figure}[tbh!]
  \centering

  \begin{subfigure}{0.5\linewidth}
    \includegraphics[width=\linewidth]{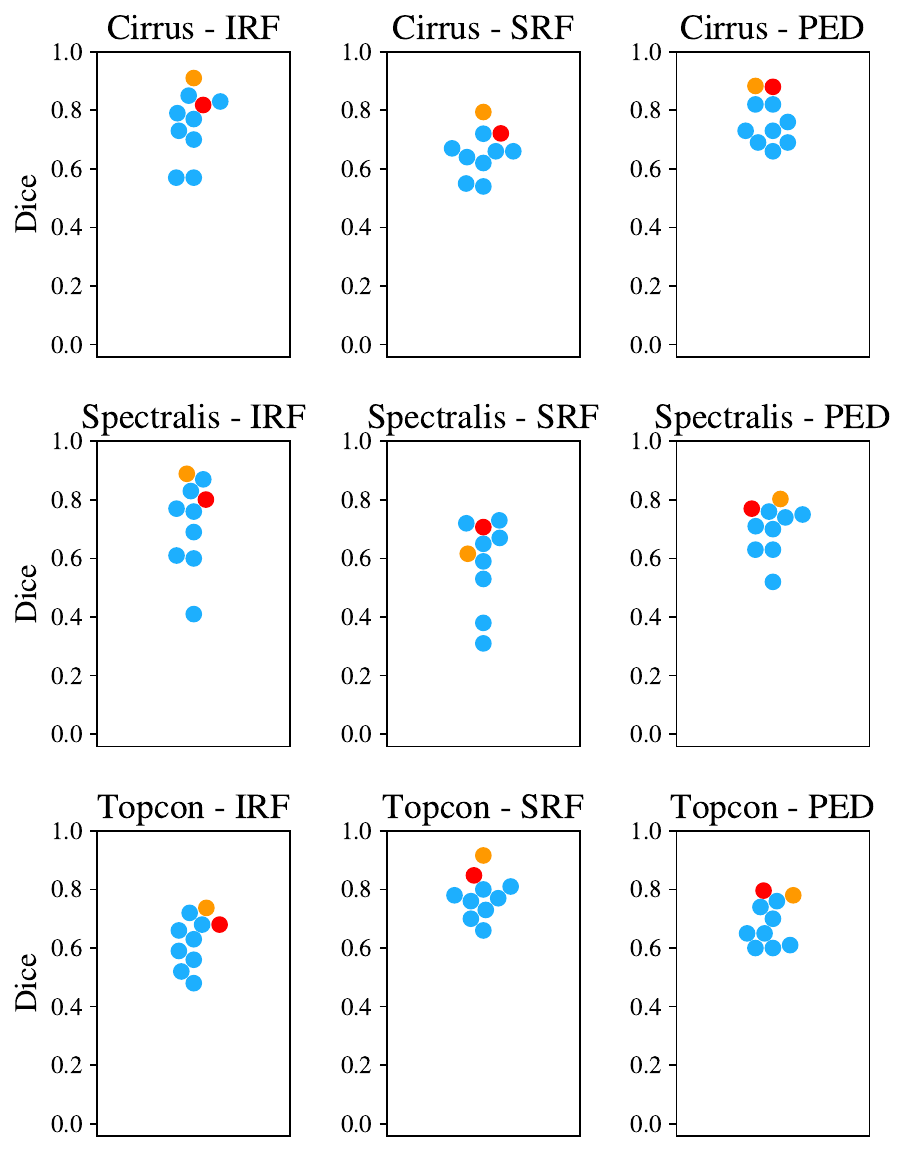}
    \caption{Dice $\uparrow$}
    \label{fig:swarm_dice}
  \end{subfigure}%
  \hfill
  \begin{subfigure}{0.5\linewidth}
    \includegraphics[width=\linewidth]{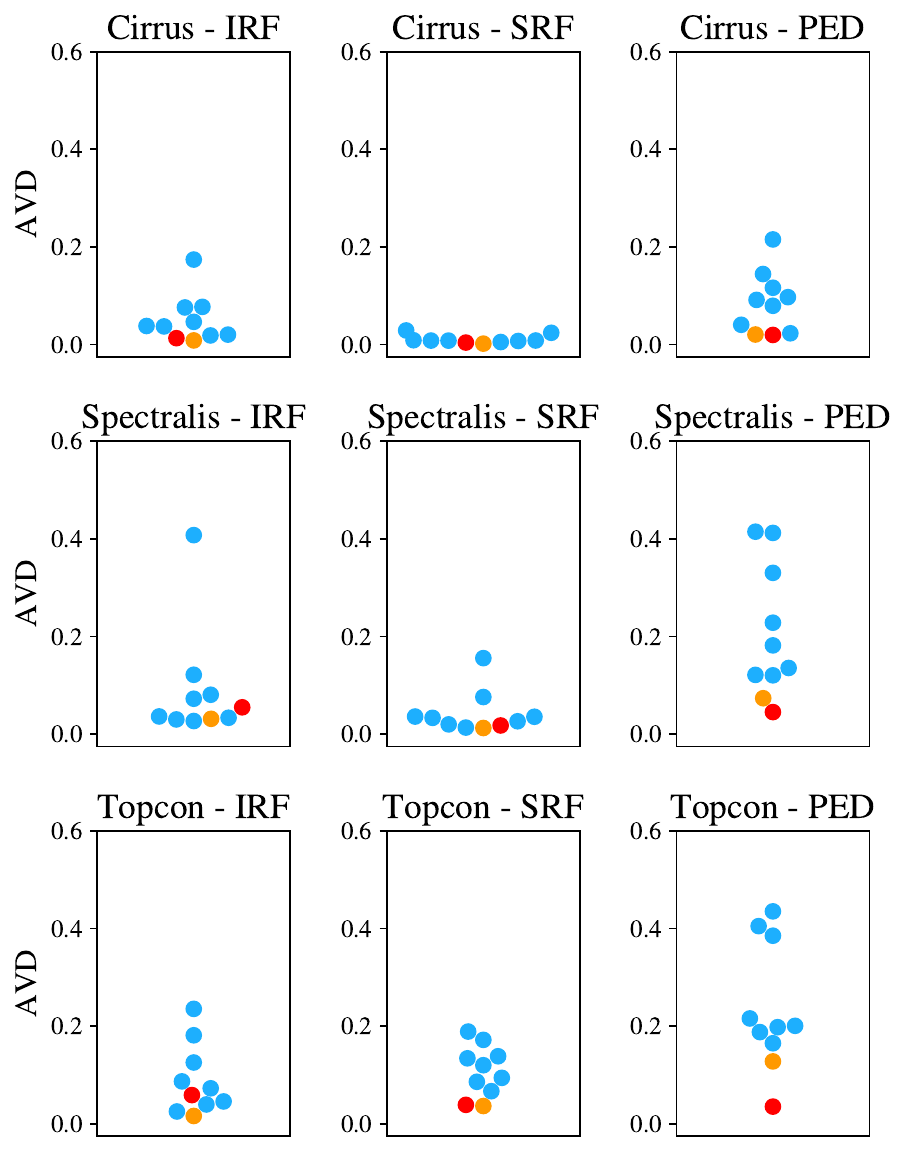}
    \caption{AVD $\downarrow$}
    \label{fig:swarm_avd}
  \end{subfigure}%
  
  \caption{Swarmplots of the detailed results per device vendor and fluid compartment. The SAMedOCT model (red) performed worse than the best-performing baseline method (nnU-net, orange) but outperformed most of the other baseline methods (challenge participants, blue) both in terms of Dice and in AVD. The challenge participants were (in alphabetical order): Helios~\cite{yadav2017generalized}, MABIC~\cite{kang2017deep}, NJUST~\cite{chen2017automatic}, RetinAI~\cite{apostolopoulos2017simultaneous}, RMIT~\cite{Tennakoon2018}, SFU~\cite{lu2019deep}, UCF~\cite{morley2017simultaneous} and UMN~\cite{rashno2017detection}.}
  \label{fig:detailed_results}
\end{figure}

\begin{figure}[tb!]
  \centering
\includegraphics[width=\linewidth]{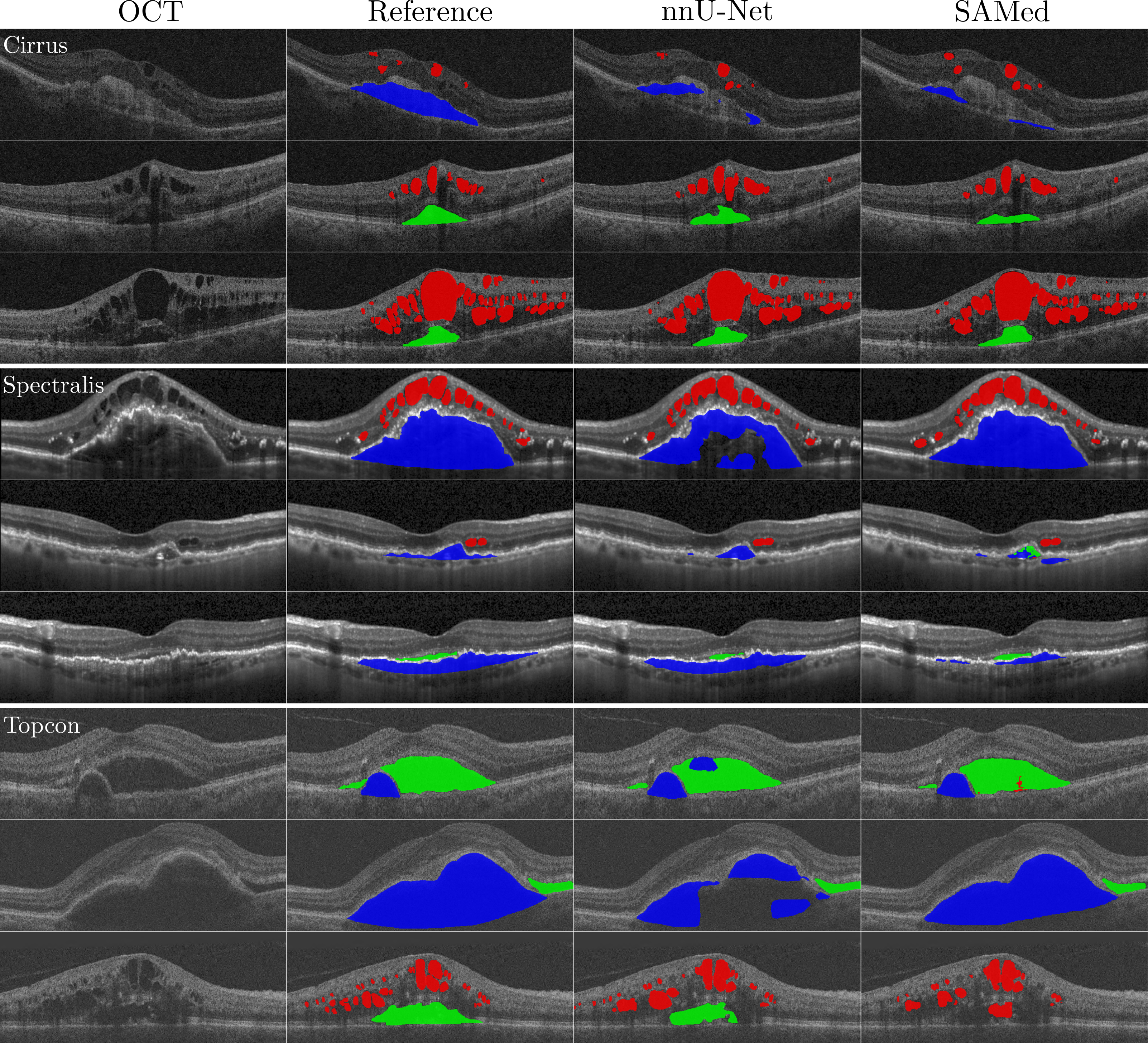}
  \caption{Qualitative results grouped by OCT vendors. Columns (left to right): Sample B-Scan from the test set, manual reference segmentation, nnU-Net segmentation, SAMedOCT segmentation. The fluid compartments displayed are: IRF (red), SRF (green) and PED (blue).}
  \label{fig:qualitative_examples}
\end{figure}

As shown in Table~\ref{tab:ablation}, the zero-shot performance of SAM is in general worse than those of the methods at least partially fine-tuned on the dataset. This aligns with findings from various studies conducted on different medical imaging modalities \cite{ji2023sam,huang2023segment,deng2023segment,ji2023segment,roy2023midl}. When examining SAM's performance with varying numbers of prompts, we observe a slight advantage for 3-point prompts over 1-point prompts, while 10-point prompts demonstrate a notable enhancement compared to the 3-point prompts. Fine-tuning the decoder led to a substantial performance improvement, with an increase in Dice score of up to 50\% compared to zero-short segmentation, however it lags behind the performance of the SAMedOCT model with trained LoRA adapters.

The detailed results per device vendor and fluid compartment (Fig.~\ref{fig:detailed_results}) indicate that the SAMedOCT model outperforms most of the baseline methods and would have been the winner of the 2017 RETOUCH challenge. However it performs worse than the nnU-net model in most of the device/disease configurations. Interestingly, the SAMedOCT model demonstrates state-of-the-art performance on PED (Mean AVD SAMedOCT: 0.033, nnU-net: 0.073). Likely due to the prominence of PED as typically the largest fluid compartment in the retina, which is often well-demarcated, it aligns more closely with the original SAM training set.

The qualitative results (Fig.~\ref{fig:qualitative_examples}), unveil both limitations and strengths of the SAMedOCT method. While it excels in correctly segmenting serous PEDs even in challenging cases where Bruch's membrane is partially obscured (Rows 4, 8), it struggles with smaller fibrous PED cases (Rows 1, 5-6). Additionally, it occasionally produces anatomically implausible predictions for IRF (Rows 7, 9).

\section{Conclusion}
This study assesses the applicability of SAM for biomarker segmentation in retinal OCT. Our results indicate that, despite achieving competitive performance, the adapted SAM slightly underperforms in comparison to networks specifically designed for medical image analysis. Nevertheless, this is still remarkable given that the SAM encoder was trained on natural images only. Further self-supervised fine-tuning of the encoder on OCT images is expected to boost the SAM performance even further, possibly beyond the one achieved with nnU-net. 
Finally, the semi-interactive nature of SAM makes it a particularly attractive approach in the clinical setting, as well as for semi-automated annotation procedures, as it allows to adjust segmentations in complex pathomorphological manifestations and account for user subjectivity.

\section*{Acknowledgements}
The financial support by the Christian Doppler Research Association, Austrian Federal Ministry for Digital and Economic Affairs, the National Foundation for Research, Technology and Development is gratefully acknowledged.

%
%
%
\bibliographystyle{splncs04}
\bibliography{paper23}

\end{document}